\documentstyle[twocolumn,prc,aps]{revtex}
\input epsf
\newcommand{\be}{\begin{eqnarray}}
\newcommand{\ee}{\end{eqnarray}}

\begin{document}

\twocolumn [ \hsize\textwidth\columnwidth\hsize\csname @twocolumnfalse\endcsname
  \title {
  Probing the Boundary of the Non-Perturbative QCD\\ 
 by  Small Size Instantons
  }
  \author {
     E.V.~Shuryak
  }
  \address {
     Department of Physics and Astronomy, State University of New York, 
     Stony Brook, NY 11794-3800 
  }
  \date{\today}
  \maketitle    

\begin{abstract}
The lattice data on the instanton size distribution
suggest an additional O($\rho^2)$ action.  The small-$\rho$
effect O($\rho^4)$ predicted by the Operator
Product Expansion (OPE) is not observed. 
 Similar deviations are also found for
non-perturbative response to a small static color dipoles: it is O(r)
not O(r$^2$).  We suggest that small instanton radii in the QCD vacuum 
and small radii of the QCD strings (to which this observation relates)
are consequences of the same phenomenon: a very
robust dual superconductivity in the QCD vacuum, with relatively large
Higgs VEV and surprisingly large Higgs mass. 
\end{abstract}

\vspace{0.1in}

]

\begin{narrowtext}   
\newpage 
1.While perturbative treatment points toward the $\Lambda_{QCD}\sim
200 \ MeV$  as 
the momentum scale where it becomes inapplicable, it is well known
by now that
the non-perturbative
phenomena actually  turn on at larger momenta. Where it happens
depends on a particular physical problem considered:
 at least {\em three different scales} have been identified so far.

 The first is the so called
{\em chiral scale}
$\Lambda_\chi\sim 1 \ GeV$, the  upper limit of
 low energy effective
theories
such as chiral effective Lagrangians or
Nambu-Jona-Lasinio model \cite{NJL} (and  it is thus
the oldest one,
identified already in  1961). Its other incarnation is a 
momentum scale at which
QCD sum rules for scalar and pseudoscalar channels fail \cite{NVSZ}.

The second scale, identified two decades later  \cite{NVSZ},
 is larger $\Lambda_{0^{\pm}glueballs} \sim (3-4)  \ GeV$. It is defined as momenta at which
 spin zero $gluonic$ correlation functions (of operators
 $G_{\mu\nu}^2$ and  $G_{\mu\nu}\tilde G_{\mu\nu}$) deviate from their
 perturbative behavior. (Note: it is $not$ the glueball masses!)

The physical origin of both these scales has been
traced down to {\em instantons-induced effects} \cite{Shu_82}. For recent detailed
review of these issues see
\cite{SS_98}, for more recent comparison between the QCD and the Seiberg-Witten
solution for the N=2
supersymmetric theory see \cite{RRS}.

2.This letter is devoted to the {\em third}  non-perturbative
scale,  related with the onset of
 confinement forces. We use instantons again, 
but only  as a small probe: 
 other small probes should show similar effects.
   Thinking about confining forces at $small$ distances
 may appear strange: it is indeed true that their
   manifestation at large distances is more important. Nevertheless,
 we study the non-perturbative corrections to properties of small-size
 ``color dipoles'', of three different kinds.

Historically the first example
is states of  heavy quarkonia.  The non-perturbative correction to
their energies  
was calculated by Voloshin and Leutwyler \cite{VL} by OPE:
$\delta E \sim  <0|G_{\mu\nu}^2|0>
r^2 \ \tau$ where the spatial size $r\sim 1/(\alpha_s M)$ and the
rotational time $\tau \sim 1/(\alpha_s^2 M)$ for large
quark mass M and small  $\alpha_s$.
I am not aware of any
precision studies of whether
 in this case it indeed works.

 Instantons is another kind of
  ``dynamical dipoles'', now with $r\sim \tau\sim\rho$.
In fact,
they are much more sensitive tool because  the probability
of the tunneling events contains the perturbative charge and all corrections 
in the exponent. As noticed in \cite{Shu_95}, it is in particular true for
 fixing the actual value of $\Lambda_{QCD}$:
  while hadronic masses and 
 quarkonium levels currently used for this purpose are
 $O(\Lambda_{QCD}$, the density of small size instantons is  $O(\Lambda_{QCD})^b$
where $b=(11N_c/3-2N_f/3)\sim 10$. (This large  power 
is nothing but the famous  one-loop beta function coefficient, $N_c,N_f$ 
 are the number of colors and flavors). So, with comparable 
 accuracy, the instanton-based determination should
potentially be 10 times more accurate!
 
 Similar to Voloshin-Leutwyler correction,
  there is the  OPE-based result \cite{SVZ} predicts the
 following correction to the density of instantons of size $\rho$:
\be 
dn(\rho)=dn_{pert} (\rho)(1+ {\pi^4\rho^4 \over 2 g^4} <0|G_{\mu\nu}^2|0>+...) 
\label{e_cdg}
\ee
Note the generic  4-th power of  $\rho$:
in QCD it is the dimension of the lowest gauge invariant local
operator.
 Note also the sign: it is nothing
else but a
generic attraction resulting from $any$ second order perturbation. 
However both these conclusions happen to be
in apparent conflict with  the lattice data (see below).

Furthermore,
 recent   studies of the vacuum reaction to small-size $static$ dipoles
(see point 6 below)
show similar deviations from the OPE-based
expectations. We argue 
 that
$both$ phenomena can be naturally understood,
provided: (i) there is rather high non-perturbative scale, so that 
(ii)  one can  
 use an  effective theory rather than QCD itself; (iii) which should
include $scalar$ composite fields with a non-zero VEV.  

3. In   Fig.[1](a) we show  recent lattice data
for the  instanton size $distribution$ in pure
 SU(3) gauge theory
\cite{anna}.  (There are others but, this work  includes such refinements as
 improved
lattice action and back extrapolation to zero smoothening.)
 One can clearly see, that a rapid rise at small $\rho$
  turns into a strong suppression. 
The former behavior is consistent with the semi-classical one-loop
result \cite{tHooft}:
\be
\label{dist0}
{dN_0\over d\rho}|_{pert} = {C_{N_c} \over \rho^5}({8\pi^2 \over
  g^2(\rho)})^{2 N_c} (\rho \Lambda)^{b} 
\ee
\begin{figure}[t]
  \epsfxsize=3.5in
  \vspace{-.1in}
  \centerline{\epsffile{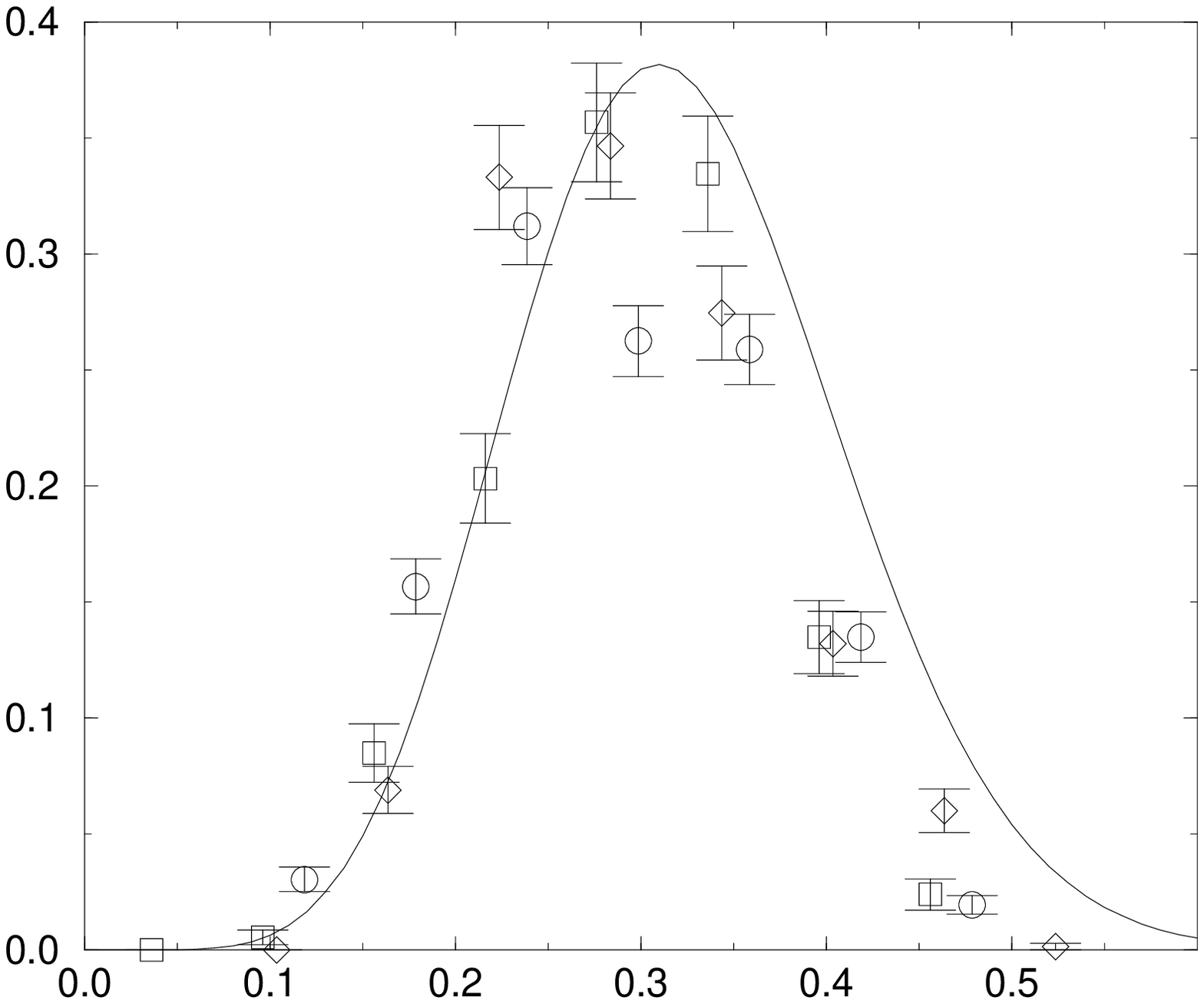}}
  \epsfxsize=3.5in
  \centerline{\epsffile{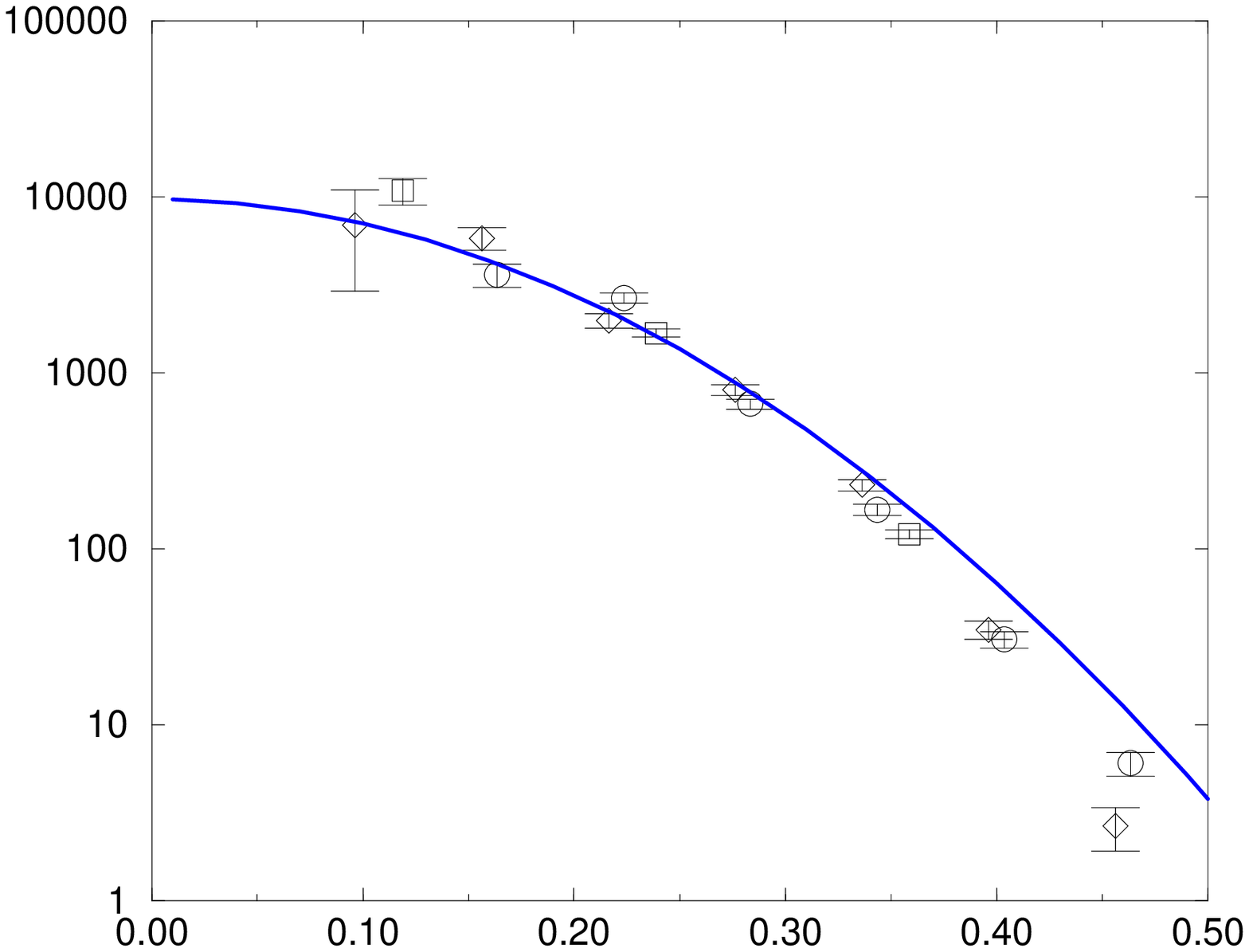}}
  \vspace{-.05in}
  \caption{
(a) The instanton density $dn/d\rho d^4z$, [fm$^{-5}$] versus its size
 $\rho$ [fm]. (b) The combination  $\rho^{-6} dn/d\rho d^4z$, in which
 the main one-loop behavior drops out for $N_c=3,N_f=0$.
 The points are from the lattice work \protect\cite{anna},
for this theory, with 
$\beta$=5.85 (diamonds), 6.0 (squares) and 6.1 (circles). Their
comparison should demonstrate that results are
rather lattice-independent.
The line corresponds to the proposed
expression $\sim exp(-2\pi\sigma\rho^2)$, see text.
  }
\end{figure}

where $C_{N_c}$ is the normalization constant, the $\rho^{-5}$ factor
and the term with the coupling constant comes
from the Jacobian
of the zero modes, and
$b=(11N_c/3-2N_f/3)$ as already mentioned.

  Sharp maximum seen in   Fig.[1](a) appears at rather small
  $<\rho>\approx 1/3$ $fm$, much smaller that  their spacing
 $R\approx 1$ $fm$.  This results in a
  non-trivial  ``vacuum diluteness'' parameter  \cite{Shu_82}
\be (\rho/R)^4 \sim (1/3)^4
\label{diluteness}
\ee  
which is so
instrumental  in understanding of many 
instantons-induced effects.
However now we are not interested in a typical instanton
but in their suppression, and so
  in  Fig.[1](b) we re-plot the same data, with the leading
semi-classical behavior
taken away. One can see the same suppression
pattern  at both sides of the maximum.  
The OPE prediction (\ref{e_cdg}) is not seen: probably it is only true
at smaller $\rho$.
The  effect is clearly
$O(\rho^2)$, and not just for small $\rho$ but in the whole region.

4. What can be the mechanism of such $O(\rho^2)$ suppression?
Before we try to answer, let us recall other suggestions from literature.
 
Diakonov and Petrov \cite{DP_84} studied the
instanton ensemble using the simplest ``sum ansatz''
for gauge field configurations and found strong
repulsion of $\bar I I$ and $I I$. It can generate 
diluteness even stronger than (\ref{diluteness}), but the result is not robust,
other trial functions have different amount of repulsion, with 
 the so called
``streamline'' configurations \cite{streamline} 
 having no repulsion for some relative
orientations. Furthermore, in dilute instanton ensemble
the suppression is different at both sides of the maximum, with
 the OPE result (\ref{e_cdg}) true
for smaller  $\rho$.

In  \cite{Shu_95} it was suggested that  more $slowly$ running
coupling constant at large distances plus
  the Jacobian-related factor $\rho^{-5}$ in the
instanton measure may be sufficient to make the density
at least convergent at large $\rho$.
 (The motivation included
 the
 well known fact that for sufficiently large number 
of flavors it  ``freezes'' at  a fixed point.) 
  But this effect cannot
be there at sizes  as small as $\rho\sim 0.1 \ fm$, and it is not
leading to the $exp(-O(\rho^2))$ law.

5.The central idea of this Letter is  
 that  $O(\rho^2)$ suppression of instantons is due to
a   ``dual superconductivity'' \cite{dual},  a scenario in which  some
composite objects 
condense, forming the non-zero
vacuum expectation value (VEV) of the magnetically charged
 scalar field $\phi$. 
 My first (naive)
argument was that in such theory, unlike the QCD itself,
  at least there is the dimension-2 operator $|\phi|^2$. 

The composites may  be
magnetic  monopoles \cite{dual,SW}, or  P-vortices, or something else:
anyway 
one is lead to an
 incarnation of the old 
Landau-Ginzburg effective theory,  Abelian Higgs Model (AHM),
describing interaction of a
``dual photon'' and ``dual Higgs''  fields. 
AHM was
applied  to the description of 
 the QCD strings, as    
Abrikosov-Nielsson-Olesen  vortices in \cite{ANO}.

Before we go into details,
let us point out a striking similarity
between these two problems. A vortex  is the 2d topologically
non-trivial 
configuration, in which $\phi$ vanishes at the
center, the Dirac string where the dual potential is singular. 
An instanton problem is in a way the previous one squared. The 
4d  picture of the fields is like two string cross sections
in two orthogonal 2d planes. Higgs field
$\phi$ again vanishes at the
center,  because  in the singular
gauge
(the only one good for multi-instanton configurations)
 the gauge field is $A_\mu(x)^2 \sim 1/x^2$  at the origin,
acting as a centrifugal barrier.
Since ``melting'' of the dual superconductor at the center
 is not a small
modification, one generally cannot expect the OPE-type calculations to
hold.
In
both problems one has first to solve for the field and then calculate
the energy or action. Fortunately, for instantons in a  Higgsed vacuum
it was already done by 't Hooft \cite{tHooft}: for fundamentally
charged
Higgs the answer is
\be
\Delta S= 4\pi^2\rho^2 |<\Phi>|^2
\label{tHooft_corr}
\ee
 Note that it leads to the $O(\rho^2)$ suppression law
we need to explain  Fig.[1](b), and that $\Delta S$
 should not necessarily be small.
We return to speculations on the exact nature of the Higgs and the
 dual photon fields of the Landau-Ginzburg model (needed to evaluate
 the
strength of the effect) at the point 7 below.

6. Now we briefly review
  applications of the dual superconductivity idea to the QCD
 string, or  the 
ANO vortex line, done in
 a series of papers  \cite{dualmodel}.
 Among clear successes of this approach is: (i) 
prediction of  weak string-string interaction, putting it around the
boundary of type I and II superconductivity;  (ii) prediction of
a whole set of potentials other than central.
 Both agree well with available lattice data, 
for a review see \cite{Bali}.

The effective Lagrangian used in  \cite{Baker_etal} is
\be L={4\over 3} [{1\over 4} (\partial_\mu C_\nu - \partial_\nu C_\mu)^2 + \ee
$$ {1\over 2} | (\partial_\mu - ig_m C_\mu)\phi |^2 +
{\lambda \over 4} ( |\phi|^2-|\phi_0|^2)^2   ]
$$
where we have omitted interaction with quarks at the ends.
$C_\mu$ is dual color potential coupled to Higgs with magnetic
coupling $g_m=2\pi/g$.
Assuming that we are exactly at
the boundary of the type I and II superconductivity,
the masses of the Higgs and the ``dual photon'' are
 equal $M_\phi=M_C=g_m \phi_0  $. The 
(classical) string tension is directly related to Higgs VEV
\be
\sigma={4\pi\over 3} |\phi_0 |^2
\label{tension}
\ee

  Lattice studies of the QCD 
strings have shown that
   they are  surprisingly thin. According to \cite{Bali}, the
  ``energy radius'' (at which it decreases by 1/e) is about
$\delta_{1/e}\approx .18$ fm, while that for the
action distribution is about twice larger. 
 In effective dual model \cite{Baker_etal}
 the  string width  is related to masses of dual photon and Higgs,
being the {\em large non-perturbative scale} of the 3-ed kind we are speaking about.
 The data mentioned put it in the
``glueball mass  range'',  around 1.3 GeV according to  \cite{Bali}
(It is difficult for me to access the error involved.)   

This observation  also has many
phenomenological
consequences. One is just another argument explaining weak
string-string interactions known from Regge phenomenology. Another
 is ``hadron diluteness'': color  field inside hadrons
 occupy only few percent of the volume 
\be
(\delta_{1/e}/R_h)^2 \sim (1/5)^2
\label{dilute_hadron}
\ee
contrary to the MIT bag model which views
the $whole$ hadronic interior to be in  the perturbative phase.
 In other words, the  value for the bag constant  $B_{MIT}\sim 50 MeV/fm^3$
was hugely under-estimated: it is $B\sim 1000 MeV/fm^3$ or more.
(Similar but different argument was
made two decades ago in \cite{Shu_78}.)

 Another way to look at this issue is to use    
{\em small static color dipoles}, or short strings. As  time is unlimited
 there is no OPE prediction like (\ref{e_cdg}),
 but using  the second order
 dipole approximation 
\cite{static_old} instead one gets
$O(r^2)$ 
corrections to the static potential
$$
V(r) = -{4 \alpha_s(r) \over 3} { 1\over r} $$ \be
+ r^2 \int d\tau \ e^{(-{3 \alpha_s(r)\tau \over 2r})} <0|G_{\mu\nu}(\tau) U_\tau G_{\mu\nu}(0) U^+_\tau |0>)  
\ee
where the  field
strengths are separated by the time delay $\tau$, with
 $ U_\tau$  being the appropriate
parallel transports.
 However  recent  lattice data on V(r) 
 at small r \cite{Bali_small} have found 
a O(r) effect, as  suggested previously in \cite{AZ} 
\be
V(r) =- {4 \alpha_s(r) \over 3} { 1\over r} + \sigma_0 r +...  
\ee
The small-distance tension is larger than the asymptotic one
 $ \sigma_0\approx (4-5)  \sigma_\infty$, but with rather uncertain error
 from perturbative subtraction. It was shown \cite{GPZ} that
 O(r) term appears
in AHM,
 although with about the same linear potential at all r,  $ \sigma_0\approx  \sigma_\infty$.

7.  The conclusions we draw from all these observations
 are: (i)  the distances  
$r=0.1-0.2$ used in these studies are already {\em large enough} to be outside
the validity domain of 
 the  OPE; (ii) but
   an effective model  like AHM should
 rather be used, and (iii) it does provides at least qualitative explanation of
the non-perturbative effects.

Encouraged by this, we return to instantons and
 try to apply the same reasoning.
Since both  the 't Hooft correction (\ref{tHooft_corr}) and the string tension
(\ref{tension}) scales as the Higgs VEV squared, we expect qualitatively
that 
\be
\label{suppr}
{dN\over d\rho} = {dN\over d\rho}|_{pert} exp{(-C\sigma\rho^2)}
\ee
where C is some numerical constant.
 In order to find C one has to identify the scalar and the dual photon fields
of the AHM, and explain how they
are coupled to  the colored gauge field of the instanton.
In the AHM treatment of the QCD string \cite{Baker_etal}
the magnetic field of the dual photon is 
identified directly with the color-electric gauge field
inside the string. It does not create problems because this electric field can
be considered Abelian.
 
 Applying the same ideas for instantons, let us first note that their 
 $self-duality$ helps: because  electric and
magnetic fields are identical, the ``magnetic'' potential $C_\mu$ and the
original one $A_\mu$ are also the same. However both are intrinsically
non-Abelian, so only a particular component (or a combination of
those)
 can be identified with the Abelian ``dual photon''  of the effective theory.
In other words, an {\em Abelian
projection} is inevitable, and there is no unique or preferred way to do it. 
Lacking better ideas, we
 simply do what lattice people do: just select one of possible projections
and see what happens. Clearly, selecting Higgs field interacting with
a particular component of the gauge potential means breaking the gauge
group 
(which the vacuum of the Standard model does and that of QCD does not).
 But we proceed anyway, simply  re-scaling the dual fields in a way 
that their Lagrangian (1) matches the 't Hooft one. If we do so,
 it leads to identification
$<\Phi>^2=(2/3)\phi_0^2$, or the constant C in (\ref{suppr})
to be $C=2\pi$.
Putting it all together, we can now compare\cite{com3}
this result to the exponential suppression. The corresponding 
 curves are shown 
 in Fig.[1], and it works very well.

9.Brief summary. The dual superconductivity seems to be surprisingly
robust. Instanton suppression we discussed and string tension
considered before both fix the AHS Higgs VEV rather accurately.
The string size and small-r potential provide hints that
the Higgs and dual photon masses are large, in the 1-2 GeV range.
Their nature remains unclear:
 so the status of AHM Higgs  is, ironically,  not that different from
that of the Standard Model.

Outlook: one may  test our suggestions by comparing instanton suppression
in  theories
with
variable number of colors and/or flavors. 
Unfortunately available data
for the SU(2) color group or SU(3) with dynamical
fermions are not yet good enough to do so.

Another challenging set of questions is related with
instanton  suppression 
 at non-zero temperatures T and/or densities. 
 At high T or density, in the quark-gluon
plasma phase, the answer is clear: the instanton
 electric fields are again  suppressed, but now by the usual
Debye screening\cite{Shu_78}. It leads to a factor $ exp(-a(T,\mu)
\rho^2)$ 
similar to the one discussed above, where the coefficient
$a(T,\mu=0)$ was calculated in  \cite{PY} and then generalized to the
$\mu\neq 0$ in
\cite{Shu_82}. Note that confinement is not completely gone \cite{com4}.
 The most interesting point is what happens
close to the deconfinement transition. Since 2 and 3 color gouge
theories
have it of the second and the first kind, respectively, a detailed
study of the instanton suppression at $T\approx T_c$ is of great interest.

{\bf Acknowledgments}.
Let me thank A.Hasenfratz for communicating the data \cite{anna} shown
in Fig.1,
and P. van Baal and V.I.Zakharov for useful discussion.
This work is partially
supported by US DOE, by the grant No. DE-FG02-88ER40388.
%
%


\end{narrowtext}

\begin{thebibliography}{99}
\bibitem{com3} The standard string tension value
$\sigma=(450 MeV)^2$   is  traditionally used to set absolute scale
of the lattice data for non-physical theories like for 
the pure SU(3) gauge theory simulations we
used: so the instanton sizes expressed in fm is consitent with it.
\bibitem{com4}
 At high T the magnetic field is related to the 3d
gauge theory by dimensional reduction. It is well known that
3d gauge theory  confines its (spatial) Wilson loops.
 However the string tension in this regime should be
 $\sigma=O(g^4 T^2)$, and thus   this
suppression
is sub-leading to the electric one discussed in the text.
\end{thebibliography}
\end{document}